\documentclass[onecollarge]{svjour2}
\usepackage[sort&compress]{natbib}
\bibpunct{[}{]}{,}{n}{}{,} 
\smartqed  
\usepackage{epsfig,graphicx,color}
\journalname{Few-Body Systems}

\begin{document}

\title{Universality in Four-Boson Systems
}

\titlerunning{Universality in Four-Boson Systems}  

\author{T. Frederico \and A. Delfino \and M. R. Hadizadeh  \and Lauro~Tomio \and  M. T. Yamashita}

\authorrunning{Frederico et al.} 

\institute{ T. Frederico \at Instituto Tecnol\'ogico de
Aeron\'autica,
12228-900, S\~ao Jos\'e dos Campos, Brazil \\
\email{tobias@ita.br}
            \and
A. Delfino \at Instituto de F\'{\i}sica, Universidade Federal
Fluminense, 24210-346, Niter\'oi, Brazil \\
              \email{delfino@if.uff.br}           
           \and
           M. R. Hadizadeh \at
       Instituto de F\'\i sica Te\'orica, UNESP -
Universidade Estadual Paulista, 01140-070,
S\~ao Paulo, Brazil  \\
               \email{hadizade@ift.unesp.br}
             \and
             Lauro~Tomio  \at
Instituto de F\'{\i}sica, Universidade Federal Fluminense,
24210-346, Niter\'oi, Brazil \\
Instituto de F\'\i sica Te\'orica, UNESP - Universidade Estadual
Paulista,
01156-970, S\~ao Paulo, Brazil \\
\email{tomio@ift.unesp.br}
            \and
           M.T. Yamashita \at
       Instituto de F\'\i sica Te\'orica, UNESP -
Universidade Estadual Paulista, 01140-070,
S\~ao Paulo, Brazil  \\
               \email{yamashita@ift.unesp.br}
}
\date{Received: date / Accepted: date}
\maketitle

\begin{abstract}
{
We report recent advances on the study of universal weakly bound
four-boson states from the solutions of the Faddeev-Yakubovsky
equations with zero-range two-body interactions. In particular, we
present the correlation between the energies of successive tetramers
between two neighbor Efimov trimers and compare it to recent finite
range potential model calculations. We provide further results on
the large momentum structure of  the  tetramer wave function, where
the four-body scale, introduced in the regularization procedure of
the bound state equations in momentum space, is clearly manifested.
The results we are presenting confirm a previous conjecture on a
four-body scaling behavior, which is independent of the three-body
one. We show that the correlation between the positions of two
successive resonant four-boson recombination peaks are consistent
with recent data, as well as with recent calculations close to the
unitary limit. Systematic deviations suggest the relevance of range
corrections. } \keywords{Universality \and Four-boson systems \and
Bound states \and Contact potential}
\end{abstract}

\section{Introduction}~\label{intro}

Quantum few-body systems interacting with short-ranged forces is not
shaped only by three-body
properties~\cite{YamEPL06,scalesreview,hadi2011,hadi2012}.
Weakly-bound tetramers composed by identical bosons have a
characteristic scale, which is independent of the trimer one, for
resonant pairwise interaction in the unitary limit (zero two-body
binding $E_2=0$ or infinite scattering length $a$). Such property
can be revealed if one considers the general case, not constrained
by some specific strong short-range interaction. The existence of
limit cycles associated with the dependence of tetramer properties
on the the four-body scale \cite{hadi2011,hadi2012} are reported.
The calculations with Faddeev-Yakubovsky (FY) equations for a
renormalized zero-range two-body interaction demands the
introduction of two independent regularization scales. The debate on
the theoretical evidences of the short-range four-body scales on
real systems, as in cold atom traps, is underway (see e.g.
~\cite{meissner04,YamEPL06,05-sogo,07hammer_platter,08-fbsefimov-fedorov,09-fbsefimov-macek,stecher,morozpra,morozfbs,DelPRA10,hadi2011,javier}).

A universal scaling function relating the energies of two successive
tetramer states, $E^{(N)}_4$ and $E^{(N+1)}_4$ (where for $N = 0$ we
have the ground-state) and the corresponding trimer energy $E_3$ of
an Efimov state~\cite{efimov1970,efimov2011} was reported as a limit cycle
\cite{hadi2011} at the unitary limit. This scaling function applies
to the energies of two tetramers between two neighbor Efimov states.
Calculations for tetramer energies with different potential models,
local and nonlocal, with and without three-body forces, at the
unitary limit~\cite{stecher,stecherjpb,DelPRA10} for tetramer
states between excited Efimov trimers, are consistent with the
scaling function, in the sense that they exhibit the characteristic
dependence on the four-body scale, as the results obtained with the
zero-range potential with two independent scales in the FY
equations~\cite{hadi2011}. It is worth noting that, the $N + 1$ tetramer
state emerges from the 3+1 threshold for a universal ratio $E^{(N)}_4/E_3
=4.6$ for $E_2=0$, which does not depend on $N$.

It was reported in \cite{hadi2011} that the tetramers move as the
short-range four-body scale is changed and the existence of the new
scale can be also revealed by a resonant atom-trimer relaxation. The
resonant behavior arises when a tetramer becomes bound at the
atom-trimer scattering threshold. Furthermore, the independent
four-body scale implies in a family of Tjon lines in the general
case~\cite{1975tjon}. Also, the positions of two successive resonant four-boson
recombination peaks, show the effect of  a four-boson short-range
scale through a universal correlation, in a form of a scaling
function, which is relevant for actual cold atom experiments and in
fairly agreement with recent data \cite{BerPRL11,FerFBS11} and
calculations with $s-$wave separable potentials \cite{DelPRA12}, as
will be presented in an updated plot. We will discuss the systematic
deviations from this correlation, by other models and data,
suggesting that range corrections is of some importance and it is of
interest for future research.

Furthermore, we will present to some extend the momentum-space
structure of the FY components of weakly-bound tetramers  at the
unitary limit \cite{hadi2012}. We show that both channels of the FY
decomposition [trimer plus atom (T+A), or $K-$type and dimer plus
dimer (D+D), or $H-$type] present high momentum tails, which
reflects the short-range four-body scale. We also point out the
relevance of the $H$ component to bring the dependence of the
four-body short-range scale to the tetramer properties.

Our analysis gives further theoretical support in favor of the
independent role of a four-body scale near a Feshbach resonance with
implications for cold atom physics. The relevance of our study is
related to the experimental possibilities to explore universal
few-body properties with tunable two-body interactions.

In Sect.~\ref{sec:2}, we present the homogeneous set of coupled FY equations
for the zero-range interaction, properly regularized to permit to
calculate the tetramer bound states with a fixed trimer spectrum. In
Sect.~\ref{sec:3}, we discuss the concept of scaling functions for the
four-boson system, with particular attention to two of them: (i) the
correlation among the energies of successive tetramer states between
two neighbor Efimov states, and (ii) the correlation of the
position of successive peaks in the four-boson recombination. Both
scaling functions are due to the freedom in the four-body
regularization scale, and the comparison with other model
calculations shows their universal validity for short-range forces
in the Efimov limit. In Sect.~\ref{sec:4}, we present results for the
tetramer structure as given by their associated FY amplitudes.
Our main conclusions are given in Sect.~\ref{sec:5}.

\section{Subtracted FY Coupled Equations - Regularization}~\label{sec:2}

\subsection{Formalism}

The FY integral equations for the contact interaction, which has a
separable $s-$wave form, have a simplified form (see e.g.
\cite{hadi2012}). The FY components can be written in terms of
reduced functions, where the dependence on one of the relative
two-particle momentum drops out, and they depend only on the other
two independent Jacobi momentum. In the case of the $K$ components
the associated reduced ${\cal K}_{ij,k}^{\,\,\, l}$ function depends
on the relative momentum of the particle $k$, in respect to the
center of mass of the pair $(ij)$, and the on the relative momentum
of particle $l$ to the center of mass of the remaining
three-particle subsystem. In the case of the $H$ components the
associated reduced ${\cal H}_{ij, kl}$ function depends on the
relative momentum of the pair $(kl)$, and on the relative momentum
of  subsystem $(ij)$ in respect to $(kl)$.
\begin{figure}[ht]
\vspace{-0.5cm}
\begin{center}
\includegraphics[width=4in,height=2in]{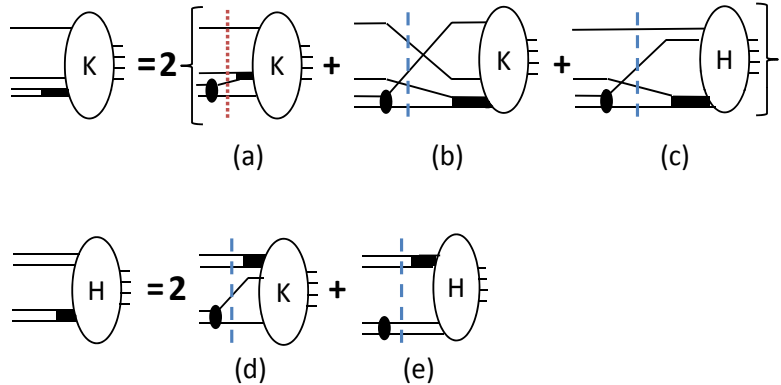}
\end{center}
\vspace{-0.2cm}
\caption{Diagrammatic representation of the
Faddeev-Yakubovsky integral equations, where the black blobs represent
two-body T-matrices for the zero-range potential (\ref{2B-T}). The
reduced FY amplitudes $\cal K$ and $\cal H$ do not depend on the
relative momentum of the pair connected by the thick black line.
{
The regularizations of the FY equations with the subtracted form
of the free four-body Green's function are indicated by the
vertical dotted and dashed lines, where the respective regulator
energy parameters are $\mu^2_3$ and $\mu^2_4$,
as shown in Eq.~(\ref{GN}).
}
} \label{FYD}
\end{figure}

The coupled set of integral equations for the reduced FY amplitudes of a
four-boson system are represented diagrammatically in Fig.~\ref{FYD}.
These integral equations have to be regularized and the four-body
intermediate state propagation should have an ultraviolet cutoff. The
kernel part labeled with (a) induces the Thomas collapse of the
{
three-body system if not regularized. One should recognize that if the
interaction between the particle $l$ and the rest of the system is turned off,
only the three-body term will survive. Then, a momentum regularization is
chosen to keep the three-boson binding energy at a given value. This
procedure is implemented by a subtraction scheme of the kernel at some
energy scale $\mu^2_3$~\cite{AdhPRL95}, which is indicated by the vertical
dotted line in the diagram. However, without regularization of the remaining
terms [(b), (c), (d) and (e)] shown in Fig.~\ref{FYD}, the tetramer will collapse,
as already discussed in Ref.~\cite{YamEPL06}~\footnote{
The occurrence of the four-body collapse can be avoided by different approaches.
One way, for example, is by regularizing the remaining terms with the same
parameter considered at the three-body level. However, in this way the
possible independent behavior of the four-body scale in relation to
the three-body one is not being explored. Another way is by considering a
different (four-body) scaling parameter (or by parameterizing the four-body
interaction with a new scale), such that a possible independent
behavior can be manifested in relation to the three-body one.
This second procedure has to be done with care when considering local
interactions in order to avoid killing any possible independent behavior
by introducing some strong short range repulsion, as it happens naturally in
the specific case of nuclear forces. As observed by Tjon~\cite{1975tjon}, in
nuclear physics the sensitivity on the number of short range scales stops at
the three-body level, because the nuclear interaction is dominated by strong
repulsive two-body forces, evidenced by the correlation between the triton
and the $^4$He binding energies when considering local two-body interactions
- the well-known {\it Tjon-line}.}.
Therefore, another subtraction is introduced together
with the intermediate four-boson propagators in diagrams (b)-(e),
indicated by the vertical dashed lines. The new subtraction
point is in principle a free parameter and, by exploring this freedom,
the tetramer and trimer energies become independent.} While in
momentum space this procedure looks straightforward, in configuration
space a short-range four-body force will play the role to separate the
trimer and tetramer energies. Another possibility is to use different
short-ranged three-body forces chosen together with on-shell equivalent
two-body potentials, by keeping the trimer energy fixed.

The reduced FY amplitudes for the four-boson bound states are the
solutions of the set of coupled integral equations (see Fig.~\ref{FYD}) given by:
\begin{eqnarray}
|\, {\cal K}_{ij,k}^{\,\,\, l} \,\rangle &=&
2\,\tau(\epsilon_{ij,k}^{\,\,\, l})\,\, \Biggl [ {\cal
G}^{(3)}_{ij;ik} \,\, |\, {\cal K}_{ik,j}^{\,\,\, l} \,\rangle +
{\cal G}^{(4)}_{ij;ik} \,\, \biggl( |\, {\cal K}_{ik,l}^{\,\,\, j}
\,\rangle + |\, {\cal H}_{ik, jl} \,\rangle \biggr)
\Biggr] , \label{FYE1} \\
|\, {\cal H}_{ij, kl} \,\rangle &=& \tau(\epsilon_{ij,kl}) \,\,
{\cal G}^{(4)}_{ij;kl}
 \,\, \Biggl[
2\,\, |\, {\cal K}_{kl,i}^{\,\,\, j} \,\rangle+ |\, {\cal H}_{kl,
ij} \,\rangle \Biggr]. \label{FYE2}
\end{eqnarray}

The free 4-body propagators in Fig.~\ref{FYD} are subtracted, and
indicated by the dashed vertical lines. The projected 4-body free
Green's function operator ${\cal G}^{(N)}$ for $N=$ 3 or 4 are
subtracted at the energies $-\mu^2_3$ and $-\mu^2_4$:
\begin{eqnarray}
{\cal G}^{(N)}_{ij;ik}=\langle\chi_{ij}|
\frac{1}{E-H_0}-\frac{1}{-\mu_{N}^2- H_0}|\chi_{ik}\rangle
\label{GN}
\end{eqnarray}
where the form factors of the contact interaction of the pair $(ij)$
is $\langle \textbf{p}_{ij}|\chi_{ij}\rangle=1$ with
$\textbf{p}_{ij}$ the relative momentum.

In the coupled equations for the reduced FY components  the
two-boson T-matrix of the pair $(ij)$, represented by the dark blob
in Fig.~\ref{FYD}, is given by
\begin{equation}
t_{ij}(\epsilon) = |\, \chi_{ij}\,\rangle \, \tau_{ij}(\epsilon) \,
\langle \, \chi_{ij}\,|, \quad \tau_{ij}^{-1}(\epsilon) =
2\pi^{2}\left(\frac{1}{a}-\sqrt{-\epsilon}\right) \ . \label{2B-T}
\end{equation}
The energies of the virtual two-body subsystem $(ij)$ appearing as
arguments of the two-boson scattering amplitude in Eq.s~(\ref{FYE1})
and (\ref{FYE2}), are $\epsilon_{ij,k}^{\,\,\, l})$ and
$\epsilon_{ij,kl}$. The former is the energy of the virtual pair, in
the 3+1 partition, while the latter in the 2+2 partition.

{
We emphasize that ${\cal G}^{(3)}_{ij;ik}$ and ${\cal
G}^{(4)}_{ij;ik}$ are matrix elements of the free-propagators of the
four-boson systems, regularized with different momentum parameters
$\mu_{3}$ and $\mu_{4}$, respectively, as shown in Eq.~(\ref{GN}). Here, it
should be clear that, instead of regularizing the integral equations
with some given momentum cutoffs, the regularization procedure is
done by subtracting the propagator at given scales, such that one
replaces the four-boson free Green's function by their subtracted
form in the FY equations. The resolvent carrying the subtraction
energy $\mu_{3}^2$, i.e., the first term on the right-hand side
of Eq. (\ref{FYE1}), keeps the trimer properties fixed.
After a close inspection of this term, together with the left
hand side, one can realize that the homogeneous trimer equation
is recovered, if the subtraction scale is the one used originally
to regularize the zero-range model for the trimer.
The extra terms brought by the four-particle FY decomposition,
in (\ref{FYE1}) and (\ref{FYE2}), are regularized with an independent
momentum parameter $\mu_4$. Within this procedure, we are sure that
the three-body subtraction scale fixes three-body observables correctly.
Note that, if we drop the extra terms brought by the interaction with
the fourth particle (say ${\cal G}^{(4)}_{ij;ik}$), we go back to the
regularized zero-range three-boson bound state Faddeev equation. In
short, the introduction of the three-body regularization parameter
$\mu_3$ avoids the Thomas-collapse\cite{AdhPRL95}, fixing a three-body
observable, whereas the regularizing parameter $\mu_4$ fixes the
four-body scale within a more general perspective of four particle
interactions~\cite{YamEPL06}.}

\subsection{Scattering Cuts and S-matrix Poles in Four-Boson Systems}

\begin{figure}[thbp]
\begin{center}
\includegraphics[width=3.5in]{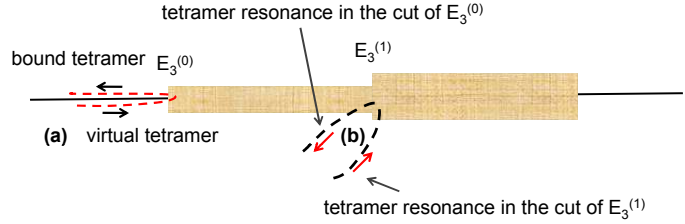}
\end{center}
\caption[dummy0]{
Schematic representation of the scattering cuts in a
four-boson system at the unitary limit in the complex energy plane. The
dashed lines with indicative arrows show different possibilities of
movement of the position of the tetramer state in the complex energy
plane when $\mu_4/\mu_3$ increases. The position of the trimer ground
and first excited states are indicated in the diagram by $E_3^{(0)}$
and $E_3^{(1)}$, respectively. In the case of a stable tetramer (a),
the arrows on the dashed line indicate the trajectory of the
corresponding state as $\mu_4/\mu_3$ increases, moving from a
virtual to a bound state.  The sketch of a possible trajectory for a
resonant tetramer state is shown in (b).} \label{traj}
\end{figure}
The variation of the subtraction parameter $\mu_4$ for fixed
$\mu_3$, moves the tetramer energies and these states can only
emerge from the  atom-trimer threshold in the unitary limit. Two
possible situations (a) and (b) are represented in Fig.~\ref{traj}
for increasing $\mu_4$ in the complex energy plane. The thresholds
and cuts associated with the atom-trimer scattering are shown in the
figure for the ground and first excited trimer.

The case (a) in Fig.~\ref{traj} corresponds to calculations
performed for stable tetramers below the trimer ground state, and
shows a possibly trajectory for the tetramer excited state coming
from a virtual four-body state. The suggestion comes in close
analogy to the behavior of an excited Efimov state, which appears
from the atom-dimer continuum threshold, as the ultraviolet
regulator parameter $\mu_3$ is increased for a fixed scattering
length \cite{YamPRA02} (see also \cite{scalesreview,TusFBS12}). Such
pattern is the same as the one already found long ago for the triton
virtual state calculated with separable nucleon-nucleon potentials,
when the deuteron binding is decreased~\cite{AdhPRC82}.
The case (b) in Fig.~\ref{traj} is an illustration of the trajectory
followed by a tetramer resonance between two successive Efimov
trimers. The resonance state emerges from the elastic cut of the
atom and first-excited trimer, passing through the threshold as the
four-body subtraction scale is increased. Then, the state becomes a
resonance entering the second energy sheet associated with the atom
and ground-trimer elastic cut. From this pattern it follows that the
tetramers close to the trimer should have a smaller width, going to
zero when the real part of the tetramer resonance energy is at the
threshold. Indeed calculations of the tetramaer resonance with
$s-$wave separable potentials shows that the tetramer closest to the
threshold has a smaller width in comparison to the other one below
it \cite{DelPRA10}. The proposed picture has yet to be checked
quantitatively within our model.

\section{Four-Boson Scaling Functions}\label{sec:3}

The four-boson observables  in the universal regime are determined
by few quantities. The physical information contained in a short
range potential can be parameterized using a two, three and
four-body scales, which corresponds to the scattering length, the
trimer bound state energy and one tetramer energy. The set of
coupled FY integral equations (\ref{FYE1})and (\ref{FYE2}) for the
zero-range interaction, is solvable once the scattering length, and
the subtraction scale $\mu_3$ and $\mu_4$ are given. Therefore, the
tetramer energies are functions of these parameters. This dependence
can be translated to a dependence on physical quantities, the
reference trimer energy and one tetramer energy, then we write that:
\begin{eqnarray}
E_4^{\rm (N+1)}=E_4^{\rm (N)}\; {\cal F}_4\left(\sqrt{E_3/E_4^{\rm
(N)}}, \pm\sqrt{E_2/E_3}\right) \ ,\label{e4}
\end{eqnarray}
in units of $\hbar=m=1$, with $m$ being the boson mass. The signs
$\pm$ distinguish between a two-body bound ($+$) or virtual ($-$)
state. The function $\cal F$ is a scaling function. On purpose we
have presented no dependence on $N$ in the scaling function, that in
principle could be the case, but it is not \cite{hadi2011}.

The use of the concept of scaling functions to study tetramer
properties comes from the successful applications to describe Efimov
trimers  in the situation where they can be bound, virtual or
resonant when the range of the interaction vanishes (see e.g.
\cite{scalesreview}). In this case only the physical scales
corresponding to energies of the dimer and a trimer are enough to
determine the successive trimer state. Remarkably, all Efimov
spectrum collapses in only one single curve, which represents the
correlation between successive trimer energies. Thus, a trimer
universal scaling function gives the energy of an excited Efimov
state, $E_3^{{\rm (N+1)}}$, as a function of $E_3^{{\rm (N)}}$,
which is written as:
\begin{equation}
E_3^{{\rm (N+1)}} = E_3^{{\rm (N)}}\; {\cal
F}_3\left(\pm\sqrt{E_2/E^{{\rm (N)}}_3}\right), \label{ev26}
\end{equation}
The universal scaling function describes all Efimov spectrum, once
the dimer and one reference trimer energy is known. It is also
useful to parameterize the Efimov excited energy showing the state
starting as virtual for $a>0$ then become bound and then becomes a
resonance at a critical value of $a<0$ \cite{resonance}. For a
finite range potential the function ${\cal F}$  exist in the limit
of $N\rightarrow\infty$ when $|a|\to \infty$, however, in practice
it converges fast for a zero-range model calculation.

The concept of scaling functions is also used to present results for
other tetramer observables, as in the case of the values of the
negative scattering length where the tetramer binding energy goes to
zero, in the Brunnian region~\cite{YamPRA10,YamFBS11}, where the Borromean
trimer has already turn into a continuum resonance \cite{resonance}.
In trapped cold atoms, both the positions or values of the
scattering length where a trimer or a tetramer cross zero,
corresponds to a resonant recombination, which produces a peak in
trap losses, and it was observed in several experiments
\cite{grimm,2009knoop,Zaccanti,2009Barontini,ferlaino2009,pollack}.
The experimental positions of resonant recombination losses, for the
trimer and the two successive shallow tetramers, were compared with
the scaling plot for the correlation between positions of successive
four boson recombination peaks in \cite{hadi2011}. In the following
we will present it compared to recent experiments \cite{BerPRL11}
and calculations \cite{DelPRA12}.

\subsection{Tetramer Energies Close to the Unitary Limit}

\begin{figure}[thbp]
\begin{center}
\includegraphics[width=5in]{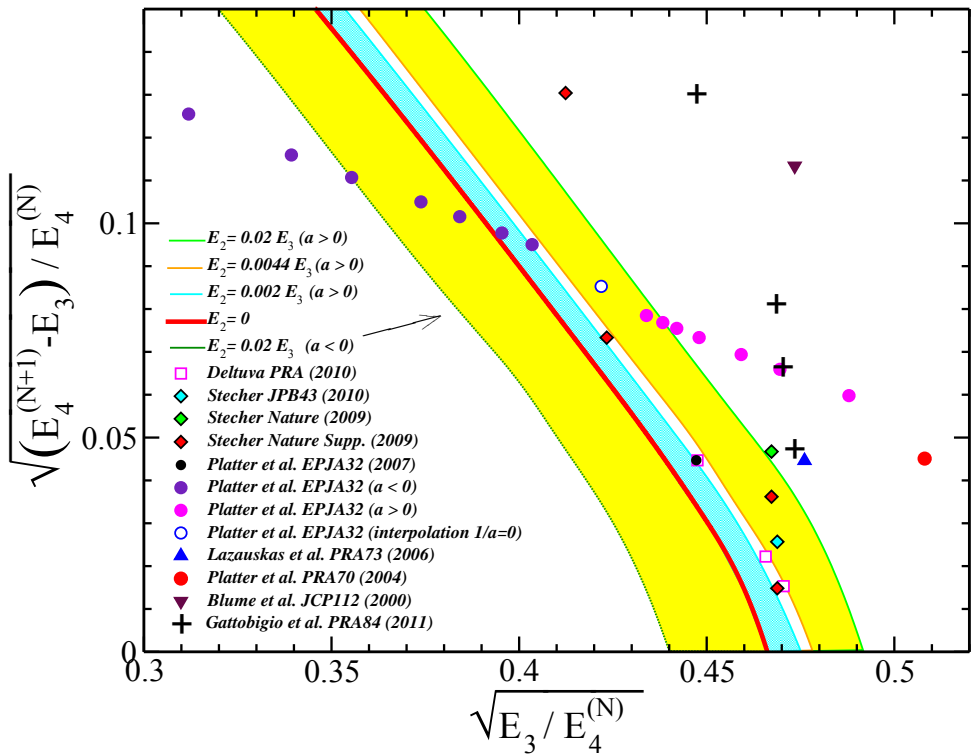}
\end{center}
\caption[dummy0]{(Color on-line)
{
The universal scaling function from Refs.~\cite{hadi2011,hadi2012}
representing the correlation of the binding energies of successive
tetramers between two neighbor trimers in a selected range. The
calculations, represented by the solid lines, were done by
considering different two-body energies, as indicated inside the
figure. The thicker line shows the unitary limiting case ($E_2=0$),
with the arrow pointing out the case with negative $a$. For
comparison, we also include other model calculations, from
Refs.~\cite{DelPRA10,stecherjpb,stecher,07hammer_platter,lazauskas06,meissner04,blume,gattobigio}
}. } \label{fig3-fbs}
\end{figure}

The correlation of the energies of successive tetramer states
appearing between two consecutive Efimov trimers in the unitary
limit is given by the scaling function evaluated at $E_2=0$:
\begin{eqnarray}
E_4^{\rm (N+1)}=E_4^{\rm (N)}\; {\cal F}_4\left(\sqrt{E_3/E_4^{\rm
(N)}}, 0\right) \ ,\label{e4ul}
\end{eqnarray}
which is presented in the scaling plot shown in Fig.~\ref{fig3-fbs}
in the form of the ratio $\sqrt{(E_4^{\rm (N+1)}-E_3)/E_4^{\rm (N)}}$.
We have performed calculations for
$N=0$ and $N=2$, i.e., up to three tetramers below the ground state
trimer \cite{hadi2011,hadi2012}. The curves for $N=0$ and $N=1$ are
not distinguishable in the scale of the figure, i.e., a limit cycle
has been achieved fast. For tetramers between two successive Efimov
trimers, the plot should be used above $\sqrt{E_3/E_4^{\rm
(N)}}>1/22.7$. With this constraint up to three tetramers between
two Efimov trimers are possible, with their energies changing
according to the value of the short-range four-boson scale,
parameterized in our model by $\mu_4$.

The validity of the scaling function to any tetramer is verified by
the consistence with other theoretical results  obtained not only
for stable tetramers but for tetramers attached to excited trimers
\cite{stecher,DelPRA10} in the unitary limit. {
This is shown in
Fig. \ref{fig3-fbs}, where the thicker solid line shows the unitary
limit (infinite $a$). We also consider corrections due to finite
scattering lengths, as indicated inside the figure, in order to verify
how sensitive are the results with respect to small positive and negative
changes of $a$.
We observe that the binding energies of excited tetramers are favored
in case of $a>0$, with the curves opening to the right side.
The same behavior is verified within effective field theory calculations
without range corrections and by considering the leading-order three-body
potential~\cite{07hammer_platter}. In this case, the unitary limit was
found by extrapolation from finite (positive and negative) values of $a$.
Other calculations with finite range potentials are shown in the figure,
where one should note that all of them are on the right side, above the
unitary limiting curve. Such results are indicative that range corrections
(which certainly are being considered in these other model calculations)
tend to open the scaling function.
This is also the case of calculations with realistic potentials for
the $^4$He trimers and tetramers performed in Ref.~\cite{gattobigio}.
}

\subsection{Position of Four-Body Resonant Losses}

The scaling function corresponding to the correlation between the
negative scattering lengths where two successive tetramers cross the
continuum threshold and produces a four-atom resonant recombination,
is written as \cite{hadi2011}:
\begin{eqnarray}
a^T_{N_3,N+1}=a^-_{N_3} {\cal A}
\left({a^T_{N_3,N}}/{a^-_{N_3}}\right) \ , \label{eq3}
\end{eqnarray}
where $a^-_{N_3}$ is the position of the three-atom resonant
recombination for $a<0$, and $a^T_{N_3,N}$ is the scattering length
for which the excited $N$-th tetramer touch the branch point of the
four-body continuum cut.

\begin{figure}[thbp]
\begin{center}
\includegraphics[width=4in]{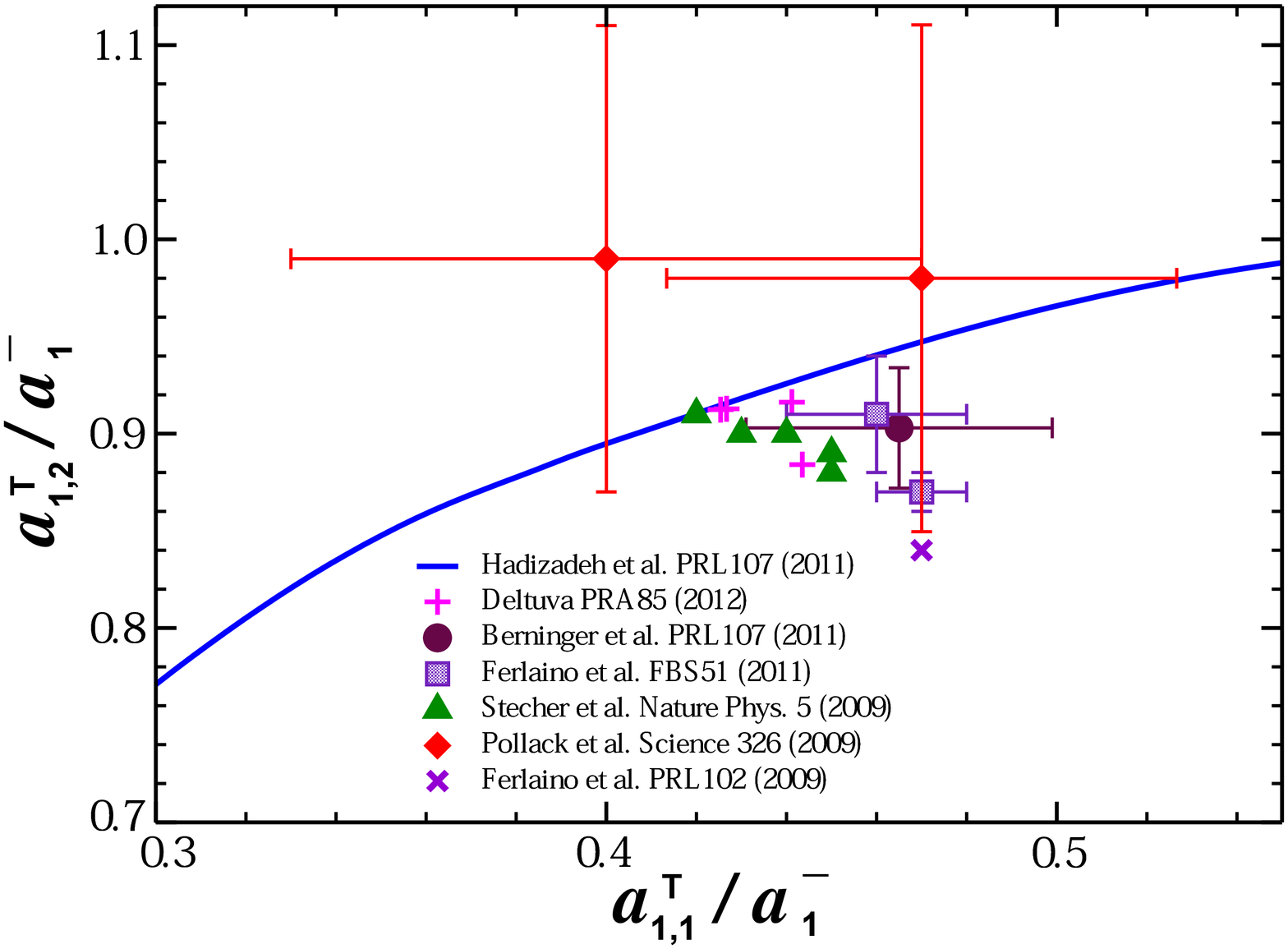}
\end{center}
\caption[dummy0]{(Color on-line)
Positions of four-atom recombination
peaks $(a<0)$ where two successive tetramers become unbound. The results
presented by the solid line, reported in Ref.~\cite{hadi2011}, are compared
with calculations done by Stecher et al.~\cite{stecher} (triangles) and
Deltuva~\cite{DelPRA12} (crosses).
The experimental results are from
Refs.~\cite{pollack,FerFBS11,ferlaino2009,BerPRL11}, as indicated in the
figure.} \label{fig4}
\end{figure}

In our calculation we have only obtained results for tetramers below
the ground state trimer, and in Fig.~\ref{fig4}, we show our
results. The scaling plot is compared to experimental data from
\cite{pollack,ferlaino2009,BerPRL11} and to other recent theoretical
results obtained for tetramers \cite{stecher,DelPRA12}, also
attached to excited trimers. The converged calculation of ref.
\cite{DelPRA12}, obtained for tetramers attached to a highly excited
trimer is agreement with the plot, while other calculations are
systematically below the curve. Such behavior suggests that range
corrections, that are certainly more relevant for the tetramers
attached to less excited trimers, may be the cause of such
systematic deviation. Range corrections will be pursued in future
works. The distortion provided by the effective range in the
universal properties of three-body systems has already been
considered within effective field theory (see e.g. \cite{PlaFBS09}).

\section{Tetramer Structure}\label{sec:4}

The dependence on the four-body scale in the solution of the FY
equations (\ref{FYE1}) and (\ref{FYE2}), comes with the coupling of
the $\cal K$ and $\cal H$ components. For instance, if $\cal H$ is
dropped out and only $\cal K$ is considered to solve the FY
equations, the dependence of the tetramer binding energy on $\mu_4$
is not relevant. Thus, the dependence on the four-body scale comes
from the coupling between the two FY components. The structure of
the tetramer, presented in terms of these components, reveals how
the dependence on the four-body scale affects both components, and
remarkably when the tetramer is strongly bound the $\cal H$
amplitude can be larger than $\cal K$ even at low momentum.
\vspace{-0.5cm}
\begin{figure}[ht]
\begin{center}
\includegraphics[width=3in]{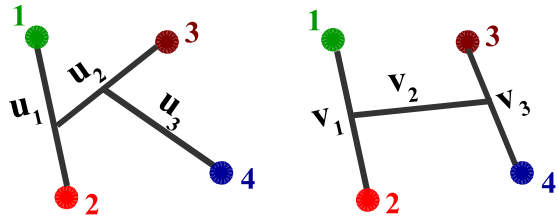}
\end{center}
\vspace{-0.5cm} \caption{
Definition of the four-body Jacobi momenta
corresponding to the $K-$ and $H-$type fragmentations.} \label{K-H}
\end{figure}
\begin{figure}[ht]
\begin{center}
\includegraphics[width=5in,height=4.3in]{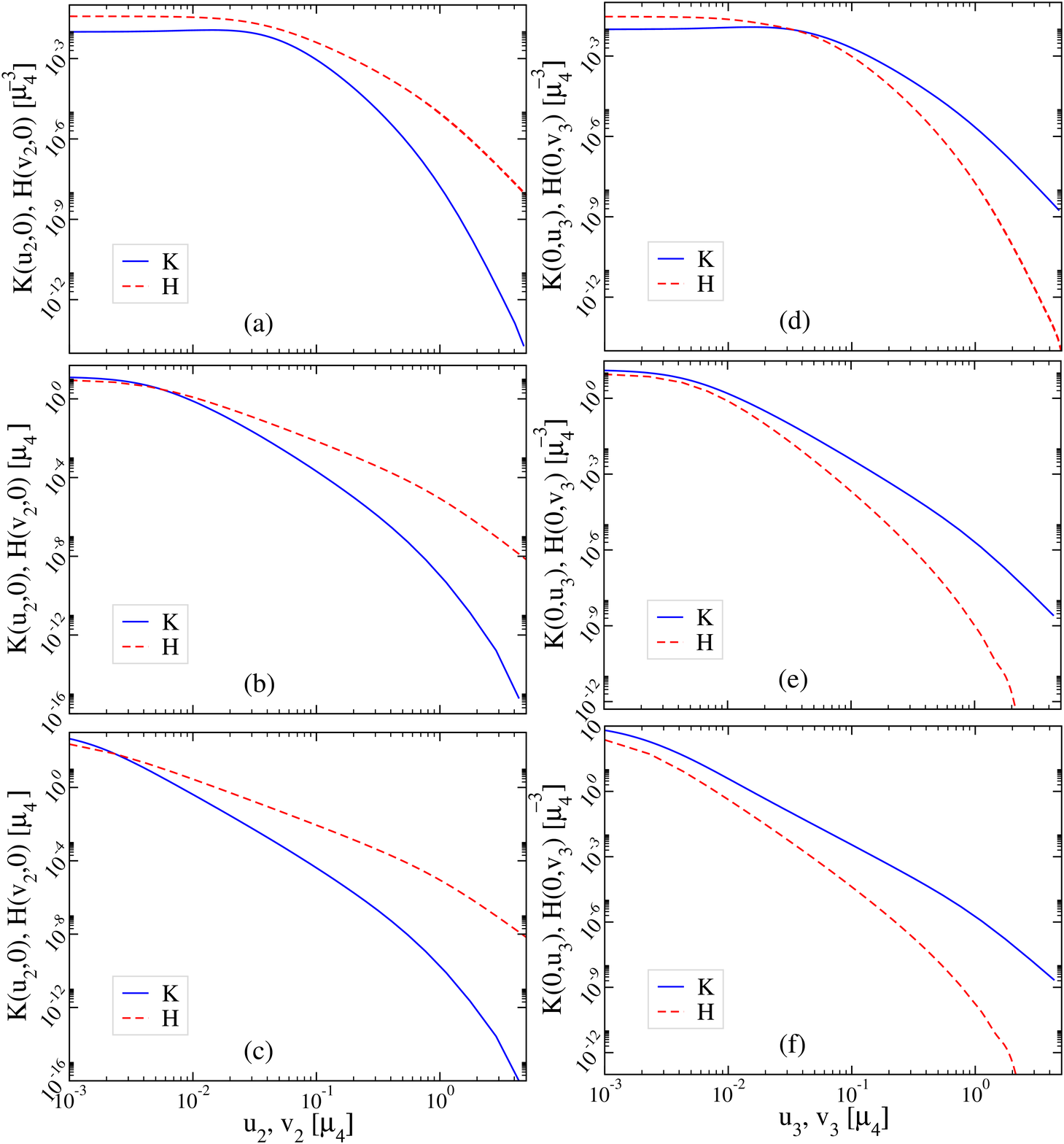}
\end{center}
\vspace{-0.5cm}
\caption{
The reduced Yakubovsky components ${\cal
K}$ and ${\cal H}$, as functions of the Jacobi momenta for scale
ratio ${\mu_4}/{\mu_3}=200$, when two four-body excited states
exist.
In the left frames we fix $(u_3, v_3)=0$; whereas in
the right frames, $(u_2, v_2)=0$. (a) and (d) are for
the ground state; (b) and (e), for the first excited state; and
(c) and (f), for the second excited state. }
\label{fig6} \vspace{-0.5cm}
\end{figure}
\begin{figure}[ht]
\begin{center}
\includegraphics[width=5in,height=2in]{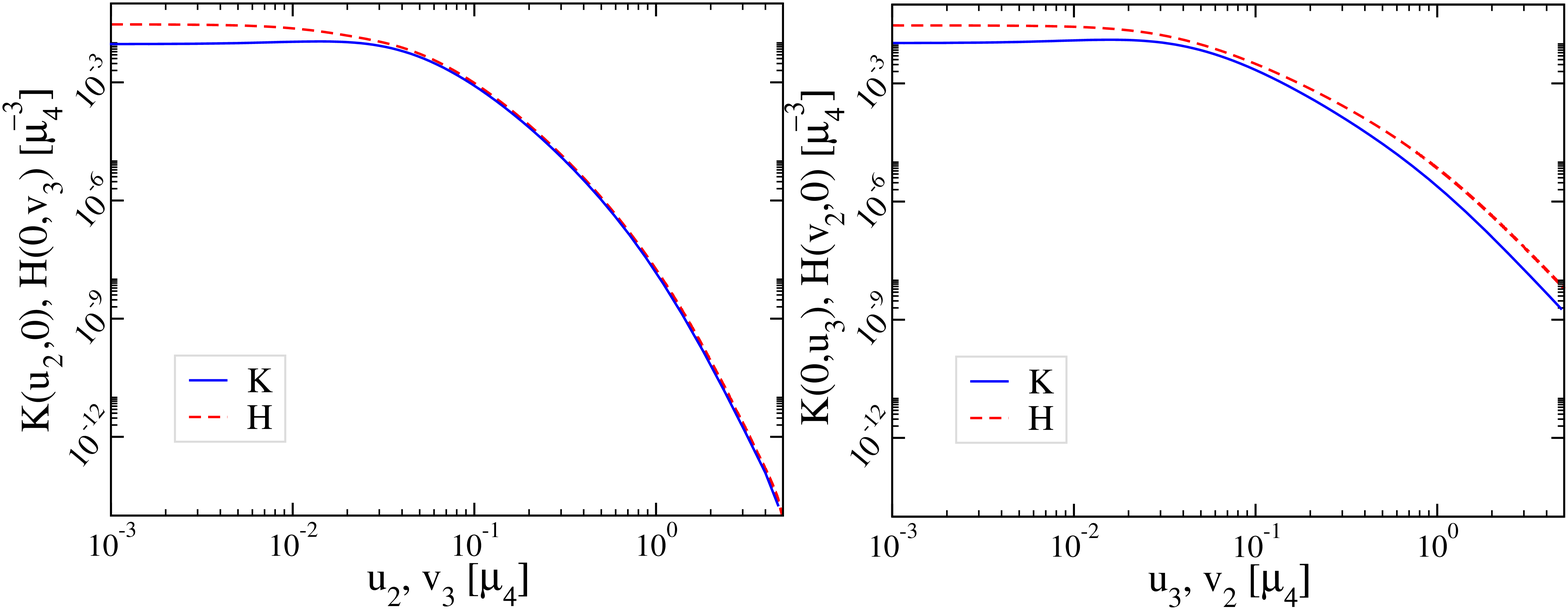}
\end{center}
\vspace{-0.5cm}
\caption{
Comparison of the reduced Yakubovsky components ${\cal K}$
(solid lines) and ${\cal H}$ (dashed lines) for the tetramer ground-state
for ${\mu_4}/{\mu_3}=200$.
From Fig.~\ref{K-H}, we should note that the Jacobi momenta directly
affected by the four-body short-range scale are $u_3$ for ${\cal K}$
and $v_2$ for ${\cal H}$. In the left
frame we are comparing the components as functions of $(u_2, v_3)$
when the ones related to the four-body scale are set to zero [$(u_3,
v_2)=0$]; and, in the right frame, we show the components as
functions of $(u_3, v_2)$ when $(u_2, v_3)=0$, in order to pin-down
the relevance of the four-body scale in ${\cal K}$ and ${\cal H}$
components. } \label{fig7} \vspace{-0.5cm}
\end{figure}

The Jacobi momenta for $K-$type and $H-$type configurations are
depicted in Fig.~\ref{K-H}. The reduced FY amplitudes have momentum
dependence given as: $\langle {\bf u}_2\; , {\bf u}_3|\, {\cal
K}_{12,3}^{\,\,\, 4} \,\rangle\equiv K({\bf u}_2,{\bf u}_3)$ and
$\langle {\bf v}_2\; , {\bf v}_3|\, {\cal H}_{12,34} \,\rangle\equiv
H({\bf v}_2,{\bf v}_3)$. The four-body regulator scale, $\mu_4$ is
carried by momenta ${\bf u}_3$ and ${\bf v}_2$ in the $K$ and $H$
configurations, respectively. Therefore, one should expect a longer
tail associated with $\mu_4 >> \mu_3$ in the momentum variable ${\bf
u}_3$ in $\cal K$ and in ${\bf v}_2$ in $\cal H$, in comparison with
the dependence in ${\bf u}_2$ and ${\bf v}_3$.  This property can be
verified by the results presented in Figs.~\ref{fig6} and \ref{fig7}
for the $s-$wave projected FY equations with a scale ratio
$\mu_4/\mu_3=200$. The results are presented by considering the normalization
such that { $\int_{0}^\infty du_2 \, u_2^2 \int_{0}^\infty du_3 \, u_3^2
\,\, {\cal K}^2(u_2,u_3)+ \int_{0}^\infty  dv_2 \, v_2^2
\int_{0}^\infty dv_3 \, v_3^2 \,\, {\cal H}^2(v_2,v_3)=1$}.
The effect of the four-body short-range scale
appears in all three tetramer states that exist for this scale
ratio, as one can verify from the results presented in
Fig.~\ref{fig6}. It is clearly observed that $K(0,u_3)$ and
$H(v_2,0)$ extend over much larger momentum than $K(u_2,0)$ and
$H(0,v_3)$, for all states. In Fig.~\ref{fig6} we choose to plot
${\cal K}$ and ${\cal H}$ in each frame by comparing the two
different momenta in respect to the dependence on the three- and
four-body short-range scales. Then one can appreciate the difference
between the momentum tails, showing the dominance of two momentum
scales well separated. The ground state results are presented in
frames  (a) and (d), the first excited state in (b) and (e), and the
second excited state in (c) and (f). Noteworthy to observe that the
$H$ reduced amplitude can be larger than the $K$ one in the low
momentum region, as shown for the case of the ground state. Even for
the shallowest and second excited tetramer state the $H$ and $K$
components are comparable as shown in frame (c). This example points
on the importance of the $H$ configuration in bringing the
dependence on the short-range four-body scale to the tetramer
physics. We remark that if the ${\cal H}$ amplitude is set to zero
in (\ref{FYE1})  and the uncoupled equation for ${\cal K}$ is
solved, the regularization parameter $\mu_4$ can be dropped off,
stressing the role of the H configuration in bringing the dependence
of the short-range four-body scale to the tetramer properties.

In particular, the expected longer tail associated with $\mu_4 >>
\mu_3$ in the momentum variable ${\bf u}_3$ in $\cal K$ and in ${\bf
v}_2$ in $\cal H$, in comparison with the dependence in ${\bf u}_2$
and ${\bf v}_3$, is further evidenced in Fig.~\ref{fig7}, which
corresponds to frames (a) and (d) of Fig.~\ref{fig6}, related to the
ground-state tetramer. As one can see from the other frames of
Fig.~\ref{fig6}, we have similar behaviors of the excited-state FY
components. The large momentum tails of $\cal K$ and $\cal H$,
manifesting the importance of the four-body scale in these
amplitudes, can be seen in the right frame of Fig.~\ref{fig7}. In
the left frame the FY reduced amplitudes drop much faster, because
these momenta are related to the three-body scale, which in this
extreme case is 200 times smaller than the four-body one. Under the
conjecture that the dependence on the four-body scale is irrelevant,
the large momentum tail of the reduced FY amplitudes should keep
only the same short-range characteristic scale, i.e., the three-body
one. Such conjecture is clearly excluded by the results displayed in
the two frames of Fig.~\ref{fig7}.

\section{Conclusions}\label{sec:5}
{
Here we report recent results on the study of the universal aspects
of four-boson systems~\cite{hadi2011,hadi2012}, as well as present
some complementary considerations. Our main remark is that the
limiting cycle of the scaling plot correlating the energies of
successive states previously found in the case of three-boson
systems close to the Efimov limit (leading to the well-known Efimov
effect), can also be verified in the general case of four-boson
systems. In particular, we  mean the correlation of the energies of
successive tetramers between two neighbor Efimov trimers. The
relevance of such results, obtained within a renormalized zero-range
model applied to Faddeev-Yakubovsky equations, is that the
four-boson scale can be driven near the Feshbach resonance by
induced four-body forces~\cite{YamEPL06}.

The scaling plot showing the universal correlation between energies
of two successive tetramer states between two Efimov
trimers~\cite{hadi2011} is compared with an updated collection of
other potential model calculations. Particularly in the present work
we include a qualitative discussion on the trajectory of the
tetramer S-matrix poles, coming out from the cuts through the
branching points, as the ratio between the three and four-body
scales is decreased.

Another scaling function, relevant for actual cold atom experiments,
correlates the positions of two successive resonant four-boson
recombination peaks~\cite{hadi2011}. It is consistent with recent
data \cite{BerPRL11} and with new calculations using $s-$wave
separable potentials \cite{DelPRA12}. Systematic deviations of
potential model results from these two scaling curves  suggest that
range corrections are of some importance and should be considered.

We have provided further insights about the dependence on the
tetramer properties on the four-body scale by analyzing the
structure of the reduced $K$ and $H$ Faddeev-Yakubovsky amplitudes
for the contact potential. It is shown that, in the reduced $H$
amplitude, the dependence on the four-body short-range scale is
evident in the behavior of the Jacobi momenta at large values
corresponding to the relative motion between two-boson subsystems in
the associated 2+2 partition. Numerically, we have confirmed that
the dependence on the short-range four-body scale can be dropped out
if one considers the solution of the uncoupled equation for the
reduced $K$ amplitude, i.e.,  setting the $H-$component to zero.

Our final remark is that a challenge remains to identify situations
in real four-boson systems where the independence between the three-
and four-body scales can be evidenced.} \vspace{-0.2cm}
\begin{acknowledgements}
We thank Funda\c c\~ao de Amparo a Pesquisa do Estado de S\~ao Paulo
and Conselho Nacional de Desenvolvimento Cient\'\i fico e
Tecnol\'ogico for partial support.
\end{acknowledgements}
\vspace{-0.7cm}


\end{document}